\documentclass[epj]{svjour}
\usepackage{graphics}
\begin{document}
\title{Phase diagram of the three-dimensional NJL model}
\author{Costas G. Strouthos
\thanks{\emph{Supported by the Leverhulme Trust.}}
}
\institute{Department of Physics, University of Wales Swansea, Singleton Park, Swansea SA2 8PP, U.K.}
\date{Received: date / Revised version: date}
%
\abstract{With the exception of confinement the three-dimensional Nambu$-$Jona-Lasinio (NJL$_3$) 
model 
incorporates many of the essential properties of QCD. We discuss the critical properties
of the model at nonzero temperature $T$ and/or nonzero chemical potential $\mu$. We show that the 
universality class of the thermal transition is that of the $d=2$ classical spin model with the same
symmetry.
We provide evidence for the existence of a tricritical point in the $(\mu,T)$ plane. 
We also discuss numerical results by Hands et al. which showed that the system is critical
for $\mu>\mu_c$ and the diquark condensate is zero.
} 
\maketitle
\section{Introduction}
\label{intro}
Phase transitions in QCD at nonzero temperature and/or 
nonzero baryon density have been studied intensively 
over the last decade both analytically and numerically.
However, since the problem of chiral symmetry breaking 
and its restoration is intrinsically non-perturbative, 
the number of available techniques is limited and most 
of our knowledge about the phenomenon comes from lattice
simulations. Because of the complexity of QCD with dynamical 
fermions, studies so far have been done on lattices with modest 
size and in various cases the results are distorted by finite
size and discretization effects.

The NJL model has been proved to 
be an interesting  and tractable laboratory to study chiral 
phase transitions both numerically by means of lattice simulations
and analytically in the form of large-$N_f$ expansions [1-9].
The lagrangian density of the $U(1)$-symmetric model is
\begin{equation}
{\cal L}= \bar{\psi}_i(\partial\hskip -.5em / + m + 
\sigma + i \gamma_5 \pi)
\psi_i
+ \frac{N_f}{2g^2} (\sigma^{2}+ \pi^2),
\label{eq:L}
\end{equation}
where the index $i$ runs over $N_f$ fermion flavors.
There are several 
features which make this model interesting for the modelling of strong
interactions:
(i) The spectrum
of excitations contains both ``baryons'' and ``mesons'',
namely the elementary fermions $f$ and
the composite $f\bar f$ states \cite{wvfn}; (ii) For sufficiently
strong coupling $g^2>g_c^2$ it exhibits spontaneous chiral symmetry breaking
implying dynamical generation of a fermion mass $m_f$,
the pion field $\pi$
being the associated Goldstone boson;
(iii) For $2<d<4$ there is
an interacting continuum limit
at a critical value of the coupling, which for $d=3$ has a numerical value
$g_c^2/a\approx1.0$ in the large-$N_f$ limit if a lattice
regularisation is employed \cite{kogut93}. There is a renormalisation group UV
fixed point at $g^2=g_c^2$, signalled by the
renormalisability of the $1/N_f$ expansion \cite{rosen91},
entirely analogous to the Wilson-Fisher fixed point
in scalar field theory; (iv) Numerical simulations
with baryon chemical potential $\mu\not=0$
show qualitatively correct behavior, in that the onset of matter occurs
for $\mu$ of the same order as
the constituent quark scale $m_f$ \cite{hands95},
rather than for $\mu\approx m_\pi/2$,
which happens in gauge theory simulations with a real measure
$\mbox{det}(M^\dagger M)$ because of the presence of
a baryonic pion in the spectrum.
This makes NJL$_3$ an ideal arena in which to test
strongly interacting thermodynamics.

In section \ref{sec:1} we discuss the universality of the 
$T \neq 0$ transition \cite{klimenko,stephanov98}.  
In section \ref{sec:2} we present results from  
a study of the phase diagram in the $(\mu,T)$ plane \cite{strouthos2001}
which support the existence of a tricritical point  on the line that 
separates the chirally broken from the chirally symmetric phases. 
We also discuss numerical results which support the non-existence
of a diquark condensate for $\mu>\mu_c$ \cite{hlm}.

\section{Universality at nonzero temperature}
\label{sec:1}
Although there is little disagreement that the chiral phase transition 
in QCD with two massless fermions is second order no quantitative
work or simulations have been done that decisively determine its universality 
class. Universality arguments are very appealing due to their beauty 
and simplicity. In essence they can be phrased as follows: At finite
$T$ phase transitions, the correlation length diverges in the transition 
region and the long range behaviour is that of the $(d-1)$ classical spin model 
with the same symmetry, because the IR region of the system is
dominated by the zero mode of the bosonic field and the contribution 
of non-zero modes does not affect the critical singularities but can be
absorbed into non-critical, non-universal aspects of the transition.
Consequently, fermions, which satisfy anti-periodic boundary conditions, 
and do not have zero modes are expected to decouple from the scalar sector. 
In another language, the classical thermal fluctuations whose energy is $O(K_B T)$ 
dominate over quantum fluctuations with energy $O(\hbar\omega)$ for soft modes of the 
field and 
the effective $(d-1)$-dimensional theory for the
bosonic fields near $T_c$ is a classical statistical theory.
A possible loophole to this standard scenario is that 
the mesons are composite $f \bar{f}$ states and their size and density 
increase as $T \rightarrow T_c$. 
Therefore, if the transition region can be described as a system of 
highly overlapping composites 
the violation of the bosonic character of the mesons may be maximal 
and the fermions become essential degrees of freedom irrespective of
how heavy they are.
\begin{table}[t]
\caption{Summary of FSS results and comparison
with $2d$ Ising and mean field scaling behaviour.}
\label{tab:1}       
\setlength{\tabcolsep}{1.3pc}
\begin{tabular}{|l|lll|}
\hline
Exponents & FSS & Z(2) & MF \\
\hline
$\nu$     & 1.00(3) & 1 & 0.5 \\
$\beta_m/\nu$ & 0.12(6) & 0.125 & 1 \\
$\gamma/\nu$ & 1.66(9) & 1.75 & 2 \\
\hline
\end{tabular}
\end{table}
At leading order in $1/N_f$ the model has a second order phase transition 
at $T_c=\frac{m_f}{2 \ln{2}}$ \cite{klimenko} with Landau-Ginzburg mean field
scaling.
We studied the critical behavior of the  $Z_2$-symmetric model with finite $N_f$
by performing lattice simulations in the vicinity of the critical point. The 
temporal lattice size was $L_t=6$ and the spatial size varied from $L_s=18$ to $50$. 
The expectation value of the auxiliary sigma field ($\Sigma \equiv \langle \sigma \rangle$)
 serves as a convenient order parameter for the
theory's critical point. 
We simulated the model exactly for $N_f=12$
with the Hybrid Monte Carlo method. The staggered fermion lattice action 
and further details concerning the algorithm can be found in \cite{kogut93}.
By using the finite size scaling (FSS) method we extracted the critical exponents. 
The results
which are summarized in Table 1 support the dimensional reduction scenario, 
because the values of the exponents are in good
agreement with those of the $2d$ Ising model rather than the mean field 
theory ones.

Next we tried to understand how the large-$N_f$ mean field theory
prediction reconciles with the dimensional reduction and universality arguments.
The answer is that the large-$N_f$
description has its applicability region. As we discuss in detail for a Yukawa theory
in \cite{stephanov98}
the phenomenon which leads to an apparent contradiction is the suppression of the
width of the non-mean field critical region by a power of $1/N_f$. 
The $2d$ Ising critical behavior sets in when $T \gg m_{\sigma}(T)$ ($m_{\sigma}(T)$ is the
thermal mass of the $\sigma$ meson).
If the cutoff $\Lambda \gg T$ the renormalized self-interaction coupling $\lambda(T)$
in the large-$N_f$ limit is close to the IR fixed point of the Yukawa theory 
and is given by
$\lambda(T) \sim T^{4-d}/N_f$ for $2\!<\!d\!<\!4$.
The mean field approximation breaks down because of self-inconsistency
when the value of the coupling of the $(d-1)$ dimensional scalar theory
$\lambda_{d-1} = T \lambda(T)$ on the scale $m_{\sigma}^{5-d}(T)$  
(the power $d-5$ comes from 
dimensional analysis) is not small anymore. Therefore, 
for $d=3$ the Ginzburg criterion for the applicability 
of the mean field scaling is given by $m_{\sigma}(T) \gg T/\sqrt{N_f}$.
This scenario was verified numerically in \cite{stephanov98}.

Additional evidence in favor of the dimensional reduction scenario 
was produced in studies of the $U(1)$-symmetric NJL$_3$ model \cite{bkt}. 
Both analytical and numerical  results showed that 
its phase structure at $T \neq 0$ is the same as the $2d$ $XY$ model.
It has two different chirally symmetric phases, one critical and one with finite
correlation length, separated by a Berezinskii-Kosterlitz-Thouless transition. 
\begin{figure}[t!]
\resizebox{0.38\textwidth}{!}{
  \includegraphics{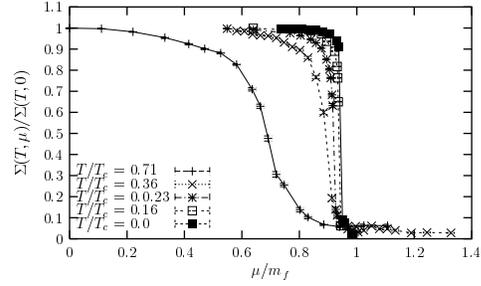}
}
\caption{$\Sigma(T,\mu)/\Sigma(T,0)$ vs. $\mu/m_f$ at different values of $T$.}
\label{fig:1}       
\end{figure}
\begin{figure}[tbp]
\resizebox{0.38\textwidth}{!}{
  \includegraphics{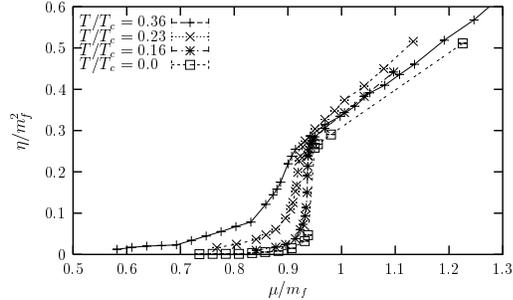}
}
\caption{$\eta(\mu)/m_f$ vs. $\mu/m_f$  at different values of $T$.}
\label{fig:2}       
\end{figure}
\section{Results at non-zero chemical potential}
\label{sec:2}
The action of the NJL model remains real even after the introduction 
of nonzero chemical potential $\mu$, which means we can study the
physics of the high density regime using standard Monte Carlo
techniques. 
In the presence of a Fermi surface with Fermi momentum $p_F$ 
the creation of $f\bar{f}$ pairs
with zero net momentum is suppressed, because one can only create
particles with $p>p_F$. So as the fermion number density $\eta(\mu)$ grows
the chiral symmetry breaking is suppressed. 
The large-$N_f$ description of the $\mu \neq 0$ chiral phase transition 
predicts a first order transition for $T=0$ and a continuous transition for
$T>0$ \cite{klimenko}. Furthermore, the critical value of the chemical potential $\mu_c$
is equal to the value of the fermion mass at $\mu=0$, which indicates that
materialization of the fermion itself drives the symmetry restoration 
transition. 
Interactions as expected decrease $\mu_c$ below the mean field result \cite{hands95}.
Work by Stephanov \cite{stephanov96} suggests that any non-zero density simulation which 
incorporates a real path integral measure proportional to $\det(MM^{\dagger})$
is doomed to failure due to the formation of a light baryonic pion from a quark $q$
and a conjugate quark $q^c$. 
The NJL model however, does not exhibit such a pathology, because the realization of the
Goldstone mechanism in this model is fundamentally different from that in QCD. 
In the NJL model the Goldstone mechanism is realized by a pseudoscalar 
channel pole formed from disconnected diagrams and the connected diagrams 
yield a bound state of mass $\approx 2 m_f$. This implies the absence 
of a light $qq^c$ state.
\begin{figure}[t!]
\resizebox{0.38\textwidth}{!}{
  \includegraphics{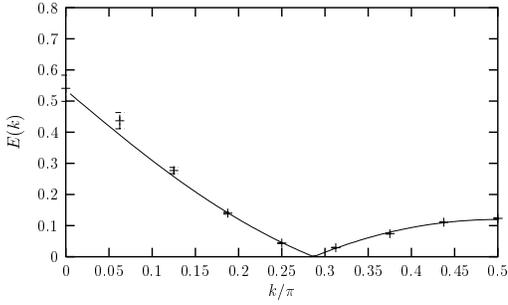}
}
\caption{Fermion dispersion relation at $\mu=0.8$.}
\label{fig:3}       
\end{figure}
As expected our simulations of the $Z_2$ symmetric NJL$_3$ 
with $N_f=4$ \cite{strouthos2001} did not provide any evidence for the 
existence of a nuclear liquid-gas transition at $\mu<\mu_c$.
It was shown in \cite{buballa} that in the NJL model there is no saturation density
for stable matter. In order to get the saturation features the authors of
\cite{buballa} introduced a chirally invariant scalar-vector interaction term which 
cures the binding problem.
Furthermore, our results showed that the second order nature of the 
$T \neq 0$, $\mu=0$ transition remains stable down to low $T$ and large
$\mu$. 
In Fig.\ref{fig:1} we plot the normalized order parameter $\Sigma(T,\mu)/\Sigma(T,0)$
as a function of $\mu/m_f$ at different values of $T$. It is clear from the shapes of these
curves that the transition becomes sharper as we decrease the temperature. In Fig.\ref{fig:2}
we plot the normalized fermion number density as a function of the chemical potential
at different values of $T$. In the limit $T \rightarrow 0$ we see that the fermion 
density is strongly suppressed before the transition and then jumps discontinuously.
By performing detailed finite size scaling analysis 
which allowed us to distinguish between second order and weak first order transitions 
we showed that the tricritical point lies on the section 
of the phase boundary defined by $T/T_c \leq0.23$, $\mu/\mu_c \geq0.97$ \cite{strouthos2001}. 
This result
shows that higher order corrections in the $1/N_f$ expansion for the nature and the location
of the transition points in the phase diagram are small in this model. 

It is well-known that at high density the diquark condensate is nonzero 
in models of strongly interacting matter which either assume that the interaction 
between quarks is due to one-gluon exchange \cite{bailin}, or by using 
effective four-fermion vertices resulting from the presence of instantons
in the QCD vacuum \cite{rapp}. Unfortunately, theoretical studies of color
superconductivity are limited to perturbative and self-consistent methods, because
of the notorious ``sign problem'' in QCD.
Hands et al. studied numerically the 
$SU(2) \otimes SU(2)$-symmetric NJL$_3$ and found no evidence for a condensate
$\langle qq \rangle \neq 0$ in the model's high density phase. 
Their results with a nonzero diquark source are consistent with a critical behavior 
$\langle qq(j) \rangle  \propto j^{\frac{1}{\delta}}$ throughout the dense phase
with $\delta$ falling in the range $3-5$ for the $\mu$ values studied. This suggests
that the model is a two-dimensional superfluid as first described by Kosterlitz and Thouless
for thin films of $^4$He
but with the universality class determined by the presence of massless relativistic fermions.
Results for the dispersion relation $E(k)$ in the spin-$\frac{1}{2}$ sector are shown in 
Fig.\ref{fig:3}.
For $k<k_F$ the lowest excitations vacate states in the Fermi sea, and hence are `hole-like', 
whereas for $k>k_F$, excitations add quarks to the system and are `particle-like'. 
There is no sign for any discontinuity on the Fermi surface characteristic 
of a BCS gap $\Delta \neq 0$ in NJL$_3$. However, recent numerical results provided evidence that 
$\langle qq \rangle \neq 0$ in NJL$_4$ \cite{walters}.
\section{Summary}
We discussed the basic features of the phase diagram of NJL$_3$.
The universality class of the $T \neq 0$ transition is that of
the $d=2$ classical spin model with the same symmetry. The non-trivial 
critical region of the $Z_2$-symmetric model is suppressed by a 
factor $1/\sqrt{N_f}$. The non-zero density phase transition is
strongly first order and the lattice simulations provided evidence for the
existence of a tricritical point on the critical line at small $T$ and 
large $\mu$. The simulations showed no evidence for the existence 
of non-zero diquark condensate at $T=0$ and $\mu>\mu_c$. The results
are consistent with a critical behaviour throughout the dense phase,
suggesting that the model is a two-dimensional relativistic superfluid.

\end{document}